\documentclass[prd, onecolumn, notitlepage, nofootinbib, floatfix]{revtex4-2}
\usepackage{amsmath}
\usepackage{graphicx}
\usepackage{dcolumn}
\usepackage{bm}
\usepackage{epsfig}
\usepackage{amssymb,latexsym,mathrsfs}
\usepackage{graphicx}
\usepackage{color}
\usepackage{hyperref}

\usepackage{float}

\hypersetup{
    colorlinks=true,
    linkcolor=red,
    citecolor=blue,
}

\newcommand{\be}{\begin{equation}}
\newcommand{\ee}{\end{equation}}
\newcommand{\bs}{\begin{split}} 
\newcommand{\bea}{\begin{eqnarray}}
\newcommand{\eea}{\end{eqnarray}}

\newcommand{\al}{\alpha} 
\newcommand{\eps}{\epsilon}

\begin{document}

\title{Logarithmic Extensions to Inflation Universality Classes} 

\author{Eric V.\ Linder$^{1,2}$} 
\affiliation{${}^1$Berkeley Center for Cosmological Physics \& Berkeley Lab, 
University of California, Berkeley, CA 94720, USA\\
${}^2$Energetic Cosmos Laboratory, Nazarbayev University, 
Nur-Sultan 010000, Kazakhstan
}

\begin{abstract} 
The values of, and connection between, the cosmological observables of the primordial power spectrum tilt 
$n_s$ and the inflationary tensor to scalar ratio $r$ are key guideposts 
to the physics of inflation. Universality classes can be defined for the tilt from the scale free 
value  proportional to $1/N$, where $N$ is the number of e-folds. 
We examine the consequences of a $\ln N$ next to leading order correction 
rather than an expansion in $1/N$, or introducing a new parameter.  
While nominally 
this can lower $r$ for  some too-high $r$ simple inflation models (e.g.\ large field 
models), there is an interesting cancellation preventing such models from 
coming back into favor.  On the other branch of the universality class, near 
Starobinsky inflation, $r$ can be raised, making it easier to detect. 
\end{abstract}

\date{\today} 

\maketitle

\section{Introduction}

Inflation -- a period of cosmic acceleration in the very early and 
energetic universe -- is on the brink of being tested in new detail 
through the detection of primordial gravitational waves. These tensor 
modes reveal the energy scale of inflation, and together with the 
scalar (density) perturbations give key indications of the inflationary 
physics. The tensor to scalar ratio $r$ and the scalar power spectrum  
slope $n_s$ form a parameter space with different classes of models 
lying in different regions of it. 

A useful and intriguing approach is to explore universality classes 
rather than individual models of the inflationary potential or slow 
roll parameters. This has the philosophy that the scalar tilt $n_s-1$ 
should be a function purely of the number of e-folds of inflation $N$, 
without additional scales entering. 

Traditionally the  ansatz has meant taking an expansion such that 
$n_s-1\sim 1/N$ plus higher order terms in powers of $1/N$. Here 
we keep the philosophy of $N$ being the determining factor, but allow 
for $\ln N$ corrections to the leading order -- still without introducing  
any other scale. We explore the effects of the next to leading order 
term, in particular on the $n_s$--$r$ 
plane, and implications for experimental limits.

\section{The $n_s$--$r$ Relation}  \label{sec:relation} 

For slow roll inflation one has a hierarchy of derivatives of the 
scalar field potential, or Hubble parameter, and from these one can 
compute the observables of the 
scalar tilt $n_s-1$ and tensor to scalar ratio $r$. 
While one can work within a specific model of the potential $V(\phi)$ 
or expansion $H(\phi)$, many classes of inflation theory exhibit a 
universality relation $n_s(N)$. One can start instead with that more 
model independent approach 
as the ansatz, as advocated early by 
\cite{1303.3925,1309.1285,1402.2059,1411.7237,1412.0678} and many 
others since then. (But see \cite{1609.04739} for 
the limitations of such an ansatz.) 

The tensor to scalar ratio $r$ is given by a differential 
equation under the slow roll assumption, 
\be 
\frac{d\ln r}{dN}-\frac{r}{8}=n_s-1\,. \label{eq:dr} 
\ee 
The standard universality relation has 
\be 
n_s-1=-\frac{\al}{N}\,, 
\ee 
where  $\al$ is of order one, and with this we can solve 
Eq.~\eqref{eq:dr} for $r$.

\subsection{$\al=$ constant} 

When $\al$ is constant, one has the well known solution 
\be 
r=\frac{8(\al-1)}{N+cN^\al}\,,  
\ee  
where $c$ is a constant of integration. This gives two asymptotic 
branches, where $r\sim N^{-1}$ and where $r\sim N^{-\al}$. 
Since under slow roll $r=16\eps$ and the slow roll parameter 
\be 
\eps=-d\ln H/dN=\frac{1}{2}\left(\frac{dV/d\phi}{V}\right)^2\,, 
\ee 
then by converting $N$ to $\phi$, 
\be 
\phi=\int dN\,\sqrt{2\eps}\,, 
\ee 
one can see that the first case corresponds to 
\be 
V(\phi)\sim \phi^{2(\al-1)}\,, 
\ee 
and the second case is 
\bea 
V(\phi)&\sim& \phi^{(2-2\al)/(2-\al)}\qquad [\al\ne2]\\ 
V(\phi)&\sim& \left(1-e^{-\phi\sqrt{2/c}}\,\right)\qquad [\al=2]\,, 
\eea 
(rolling off the nearly flat plateau at large $\phi$ in the $\al=2$ 
case).

\subsection{Logarithmic running of $\al$} 

However, $\alpha$ generally gets terms beyond the leading order 
constant term. For $\al(N)$, 
Eq.~\eqref{eq:dr} can be solved to give 
\cite{1412.0678} 
\be 
\frac{r}{8}=\left[-e^{\int (dN'/N')\,\al(N')}\,\int dN'\,e^{-\int (dN''/N'')\,\al(N'')}+c\,e^{\int (dN'/N')\,\al(N')}\right]^{-1}\,,  \label{eq:ral} 
\ee 
where $c$ is again an integration constant and the integrals are 
evaluated up to a $N$ e-folds. 

The point of universality is to not introduce any time dependence or 
scale other than the e-fold scale $N$. Thus we expect that higher order 
terms inducing a variation 
of $\al(N)$ should be in some series expansion in $N$ with coefficients 
of order one. The natural ansatz is $\al(N)=\al_0+\al_1/N+\al_2/N^2+\dots$. 
This has been studied for many cases and arises from, for example,  
hilltop inflation \cite{1309.1285}. 

One could also consider $\ln N$ effects, and indeed Starobinsky 
inflation \cite{staro} has $\al(N)=2-3\ln N/N$. Since $\ln N/N>1/N$ then 
we expect this correction to play a larger role in altering the relation 
in the $n_s$--$r$ plane than a simple $1/N$ next to leading order term. 
However, there is potentially an even larger next to leading order term: 
$1/\ln N>\ln N/N$. This is the term we consider in this paper: 
\be 
\al(N)=\al_0+\frac{\al_1}{(\ln N)^s}\,, \label{eq:lnn} 
\ee 
where the next to leading order term is suppressed by $(\ln N)^s$. 

Using Eq.~\eqref{eq:ral} this gives, for $s\ne1$, 
\be 
\frac{r}{8}=\left[-N^{\al_0}e^{[\al_1/(1-s)](\ln N)^{1-s}}\,\int dN\,N^{-\al_0} e^{-[\al_1/(1-s)](\ln N)^{1-s}}+cN^{\al_0}e^{[\al_1/(1-s)](\ln N)^{1-s}}\right]^{-1}\,. 
\ee 
The $s=1$ case is simpler, with 
\be 
\frac{r}{8}=\left[-N^{\al_0}(\ln N)^{\al_1}\,\int dN\,N^{-\al_0}(\ln N)^{-\al_1}+cN^{\al_0}(\ln N)^{\al_1}\right]^{-1}\,. 
\ee 
For simplicity, we consider two analytic cases: Case A with $s=1$ and $\al_1=-1$, and Case B 
with arbitrary $s$ and $\al_0=1$. 

Case A yields 
\be 
\frac{r}{8}=\left[\frac{N}{(\al_0-1)}+\frac{N}{(\al_0-1)^2\ln N}+\frac{cN^{\al_0}}{\ln N}\right]^{-1}\,.  
\label{eq:rcaseA} 
\ee 
This is interesting: the first branch gains an extra (positive) term 
in the denominator, lowering $r$  for a given $\al_0$, while 
the second branch has a $\ln N$ suppression, raising $r$. 

\begin{figure}[!htb]
\centering 
\includegraphics[width=0.6\columnwidth]{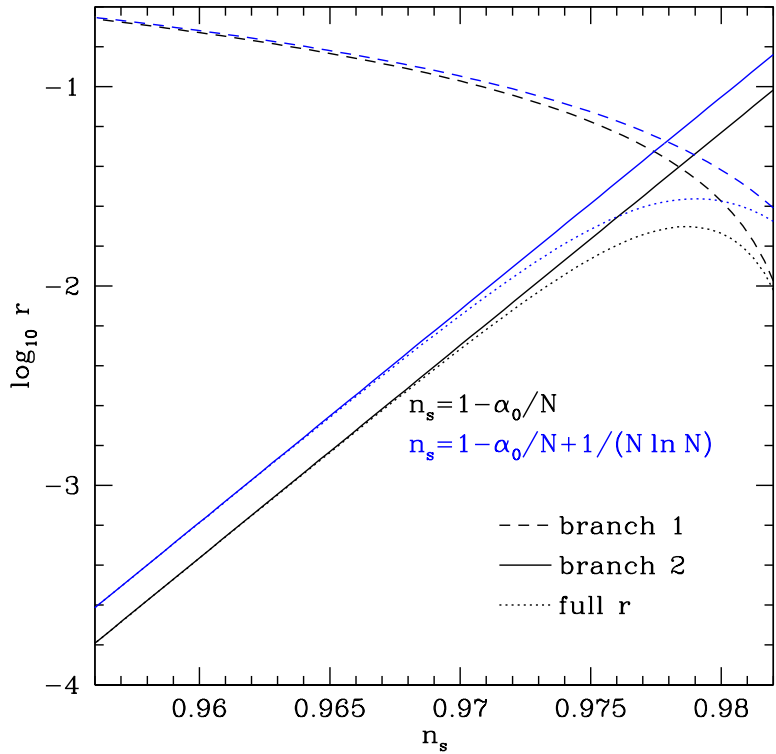} 
\caption{
Relation between $r$ and $n_s$ for inflation models following 
the standard leading order ansatz (black) or including a 
next to leading order $1/\ln N$ term (blue; Case A: $s=1$, $\al_1=-1$). 
Asymptotic branches due to the two terms are shown, dashed 
and solid (with $c=1$) respectively, while the dotted curves give 
the full sum. Curves are shown for $N=60$ 
and $\al_0$ runs from 3 to 1.05, from slightly off the left side to slightly off the 
right side.  
} 
\label{fig:rns} 
\end{figure}

Figure~\ref{fig:rns} shows the behavior in the $n_s$--$r$ plane 
for the standard relation (leading order term only), and for Case A with the 
next to leading order $1/\ln N$ term, i.e.\ Eq.~\eqref{eq:rcaseA}. In the region of greatest interest, 
$0.96\lesssim n_s\lesssim 0.97$, the second branch dominates 
for $c\sim1$ (here we took $c=1$). Indeed the next to leading order 
term enhances the value of $r$ for a given $n_s$, making 
inflationary gravitational waves easier to detect.

The next to leading order term 
gives a multiplicative enhancement of $r$ on the second branch 
by a factor $\ln N\approx4$ 
(we show results for $N=60$) for fixed $\al_0$. However, it shifts $n_s$ as well, so 
the vertical and horizon displacements combine to give a smaller 
gain in r, by $(\ln N)/e\sim1.5$. Since the first branch has shallower slope, 
the horizontal shift actually erases the expected lowering of $r$, 
giving a slight enhancement. In any case, this region of relatively 
high  $r$ is disfavored by data, especially after the new BICEP/KECK 
results, $r<0.036$ at 95\% confidence level \cite{bicep}. 

\begin{figure}[!htb]
\centering 
\includegraphics[width=0.6\columnwidth]{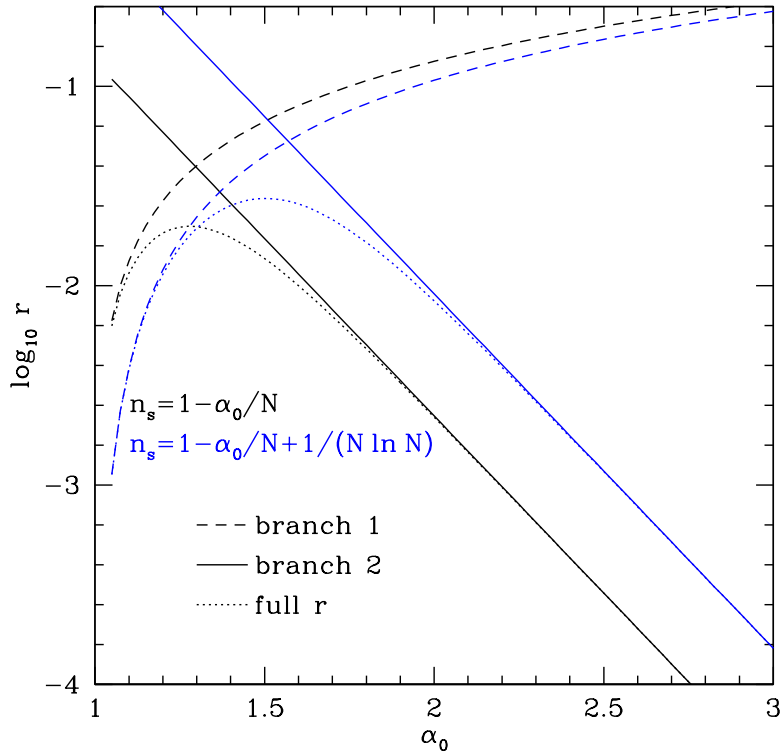} 
\caption{
As Fig.~\ref{fig:rns} for Case A but now in the $r$--$\al_0$ plane, 
so $n_s$ runs along the curves. Values $\al_0\approx2$ 
give $n_s\approx0.96$--0.97. 
} 
\label{fig:ralpha} 
\end{figure}

To clarify the shift induced in $r$ we plot the $r$--$\al_0$ plane 
for Case A in Fig.~\ref{fig:ralpha}; this gives a purer picture of $r$ but note 
that a given value of $\al_0$ will have differing values of $n_s$ 
between the curves using leading order alone and those including 
the next to leading order. The factor of four enhancement for the 
second branch and the reduction by a factor $1+1/[(\al_0-1)\ln N]$ 
for the first branch are clear.

Turning to Case B, with arbitrary $s$ but $\al_0=1$, gives 
\be 
\frac{r}{8}=\left[\frac{N(\ln N)^s}{\al_1}+cN\,e^{[\al_1/(1-s)](\ln N)^{1-s}}\right]^{-1}\,, 
\ee 
or when $s=1$,  
\be 
\frac{r}{8}=\left[\frac{N\,\ln N}{(\al_1-1)}+cN\,(\ln N)^{\al_1}\right]^{-1}\,.  \label{eq:ral0s1}
\ee 
For integer $s>1$, we find $n_s$ too high to be viable, and a 
low $r$, so the $s=1$ case is most relevant, with a large enough  
$\al_1$ to give a viable $n_s-1=-1/N-\al_1/(N\ln N)$. 

\begin{figure}[!htb]
\centering 
\includegraphics[width=0.6\columnwidth]{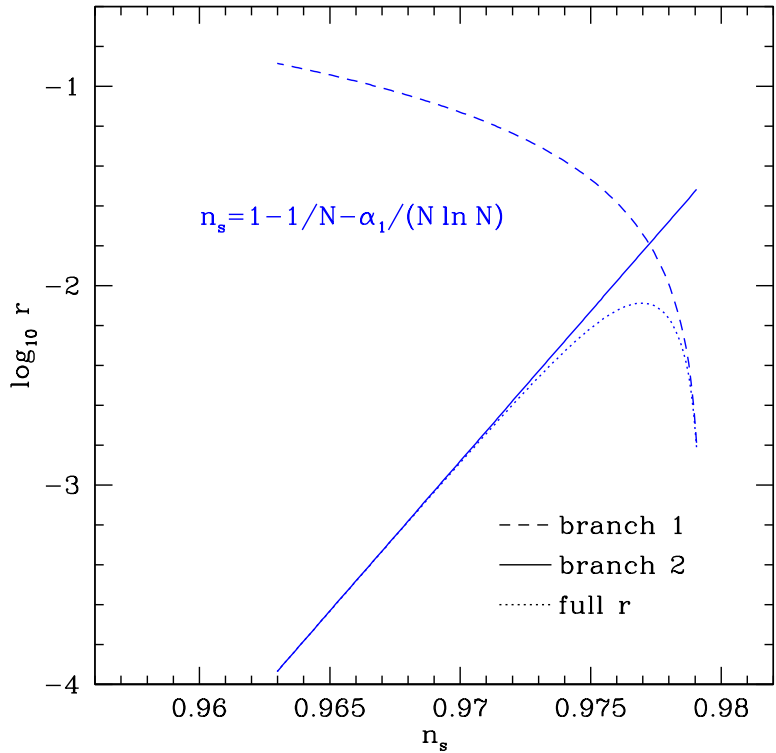} 
\caption{
As Fig.~\ref{fig:rns} but now with $\al_1$ free and fixing 
$\al_0=1$ to get a simple solution. Curves run over 
$\al_1=1.05$ (right endpoints) to 5 (left endpoints). 
} 
\label{fig:rnss1a01} 
\end{figure}

Figure~\ref{fig:rnss1a01} 
shows the results for Case B ranging over $\al_1=1.05$--5. Again the second 
branch is the dominant one for viably small $r$. In fact, 
the constraint on $n_s$ forces smaller $r$ than in the 
previous models, with $n_s=0.970$ (0.973) having 
$\al_1=3.2$ (2.5) and $r=1.4\times 10^{-3}$ ($3.6\times10^{-3}$).

\section{Conclusions} 

The ansatz $n_s-1\sim 1/N$ is an attractive quasi model independent, 
or universality class, approach to inflation. It has the virtue that no scale 
enters different from order one other than the number of e-folds of inflation. 
Beyond leading order terms that also depend only on $N$ have the same 
property. Here we examined the largest possible next to leading order 
term, $1/\ln N$. 

Logarithmic terms in the expansion also appear, though in a different form, 
in well known inflation theories such as Starobinsky inflation. Here we 
explore what impact this larger correction can have on the power spectrum 
tilt $n_s$ and tensor to scalar ratio $r$. Equation~\eqref{eq:lnn} gives 
our basic ansatz, and Eqs.~\eqref{eq:rcaseA} and \eqref{eq:ral0s1} the 
solutions of most interest. For viable values of $n_s$ and $r$, the 
second branch is most relevant,  and we find that for Case A the next 
to leading order term can make inflationary gravitational waves more 
detectable by increasing $r$ by a factor $\sim1.5$ for a given measured value of 
$n_s$.

\acknowledgments 

This work is supported in part by the Energetic Cosmos Laboratory and by the 
U.S.\ Department of Energy, Office of Science, Office of High Energy 
Physics, under contract no.~DE-AC02-05CH11231.


\end{document}